\documentclass[conference]{IEEEtran}

\usepackage{cite}

   \usepackage[pdftex]{graphicx}

   \DeclareGraphicsExtensions{.pdf,.jpeg,.png}

\usepackage{marginnote}

\usepackage{eso-pic}

\makeatother

\begin{document}

\title{A Stereolithographically Fabricated Polymethacrylate Broadband THz Absorber}

\author{\IEEEauthorblockN{Serang Park, Zackery Z. Clark, Yanzeng Li, Michael McLamb, and Tino Hofmann}
\IEEEauthorblockA{Department of Physics and Optical Science, University of North Carolina at Charlotte\\
Charlotte, North Carolina, 28223\\
Email: spark71@uncc.edu}}

\maketitle

\begin{abstract}
Additive manufactured THz optics have been introduced as an efficient alternative to their commercial counterparts. Among various additive manufacturing methods, stereolithography provides superior spatial resolution and surface finish. However, examples of stereolithographically fabricated components for THz applications are still scarce. In this paper, we report on the fabrication process and performance of a stereolithographically fabricated broadband absorber for the THz spectral range. Simple THz transmission experiments were carried out for the absorber and bulk reference samples. The experimental results indicated that the fabricated absorber effectively absorbs incident signal in the investigated THz spectral range.
\end{abstract}

\IEEEpeerreviewmaketitle

\section{Introduction}

The continuous development in optical systems for the THz spectral range demands improvements in manufacturing of the optical components. In recent years, additive manufacturing has been introduced as an effective approach for prototyping THz optical elements such as lenses, waveguides, and filters \cite{zhou2016additive,squires20153d,Weidenbach:16}. So far, the prevalent additive manufacturing technique used for the fabrication of THz optical components is fused filament deposition. This technique offers a wide range of compatible materials suitable with THz optical applications and is advantageous due to the low instrument and fabrication costs. However,  this method suffers from a low spatial resolution and surface finish \cite{Yan1996}. The spatial resolution of fused filament deposition, which is on the order of several hundreds of $\mu$m, is mainly limited by the nozzle diameter through which material filaments are heated and applied.

Stereolithography, in contrast, has been demonstrated to achieve spatial resolutions on the order of 10~$\mu$m and substantially better surface finish compared to other additive manufacturing fabrication techniques \cite{NGO2018172,Shallan2014chem}. Reflective optical components have been successfully fabricated by metalizing polymer-based, stereolithographically fabricated reflectors. Using these approaches comparable THz optical responses to commercially available optics were achieved \cite{Fullager2019, colla2019}. In addition to fabricating reflective optics, polymers compatible with stereolithographic fabrication techniques also show sufficient transparency to potentially allow the manufacturing of transmissive optics for the THz spectral range \cite{Park2019}. Thus, stereolithographic fabrication may open up new pathways for the manufacturing of polymethacrylate components for THz optical applications.

In particular, the ability to fabricate virtually arbitrary geometries to achieve THz optical properties which are difficult to be obtained using naturally occurring materials has received substantial attention in the past. Very recently, Petroff \textit{et al.} demonstrated a broadband absorber for the THz spectral range fabricated by additive manufacturing \cite{Petroff20193D}. The demonstrated absorber was composed of periodic pyramidal structures and found to be effective in this spectral range. The absorber geometries were fabricated from polylactic acid and high impact polystyrene using fused filament deposition techniques. However, the successful use of fused filament deposition techniques for these applications imposes substantial design constraints onto the target geometry. These constraints are required in order to achieve true-to-form fabrication and avoid artifacts due to limitations in filament placement and adherence. 

In this paper, we report on the fabrication process and performance of a stereolithographically fabricated broadband absorber for the THz spectral range from 82 to 125~GHz. The absorber was composed of periodic pyramidal structures arranged along a Hilbert path and fabricated from a THz transparent polymethacrylate. THz transmission measurements confirmed the effectiveness of the investigated absorber in comparison to a bulk reference sample.

\section{Experiment}

\subsection{Sample Design and Preparation}

\begin{figure}[bt]
	\centering
	\includegraphics[width=0.8\linewidth]{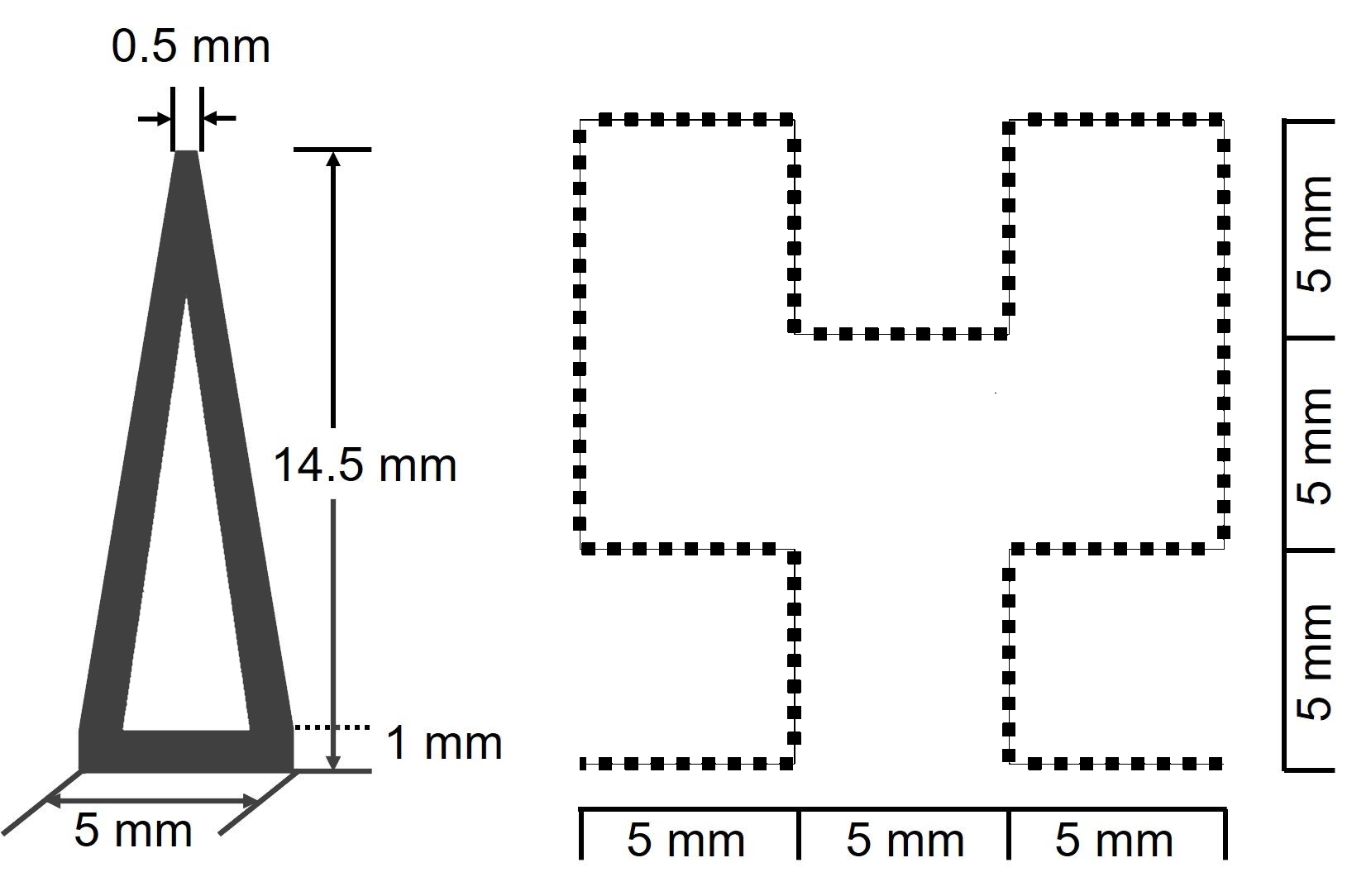}
	\caption{Triangular cross section of the absorber (left) and a unit cell of the Hilbert curve path (right). The dimensions were selected to be compatible with the commercial stereolithographic system used for the sample fabrication here.}
	\label{fig_curve}
\end{figure}

\begin{figure}[tb]
	\centering
	\includegraphics[width=0.75\linewidth, trim=0 22 0 60,clip]{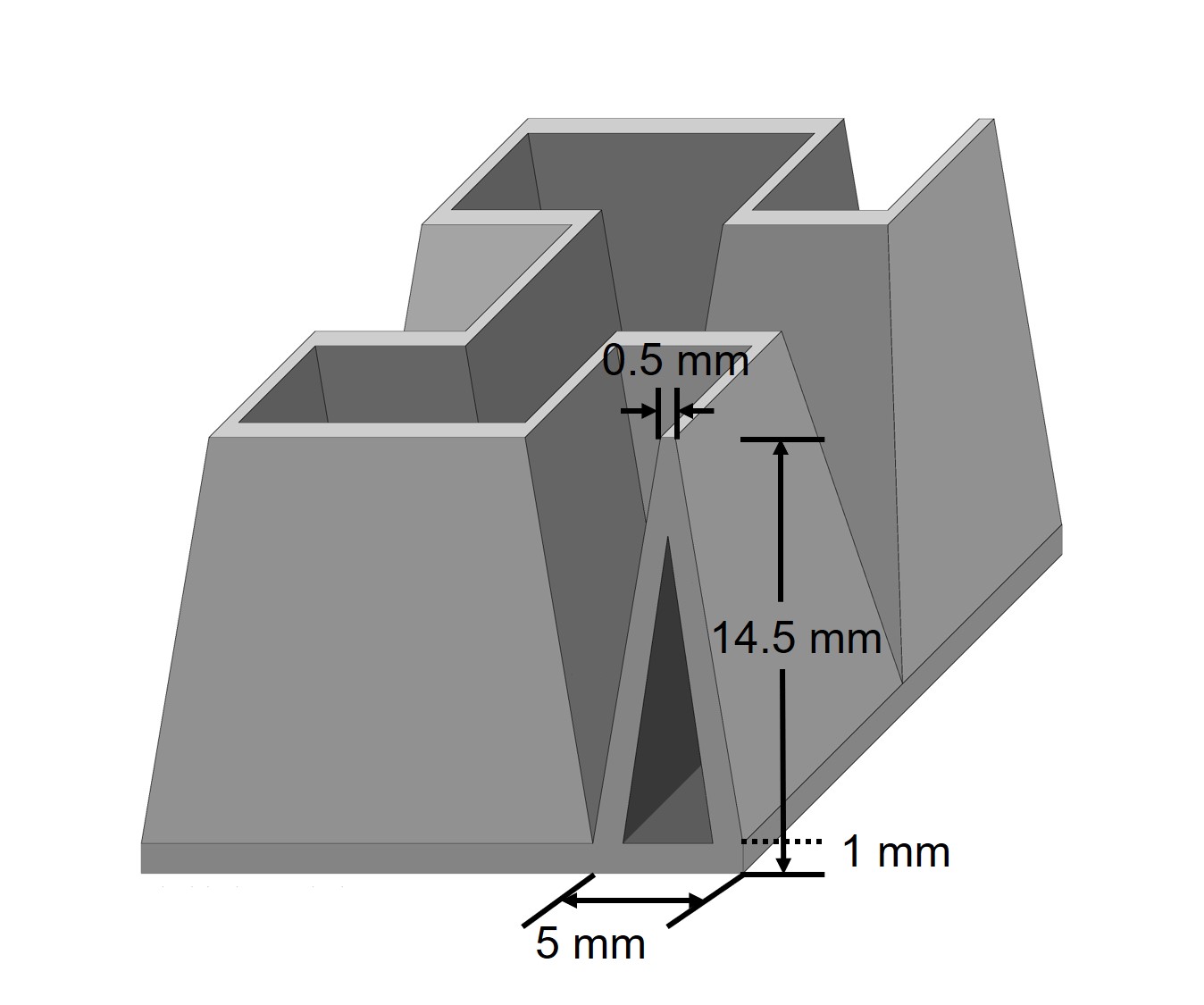}
	\caption{3D drawing of a unit cell which composes the absorber. A triangular cross section is extruded along a Hilbert curve. }
	\label{fig_3D}
\end{figure}

\begin{figure}[bt]
	\centering
	\includegraphics[width=1.9in]{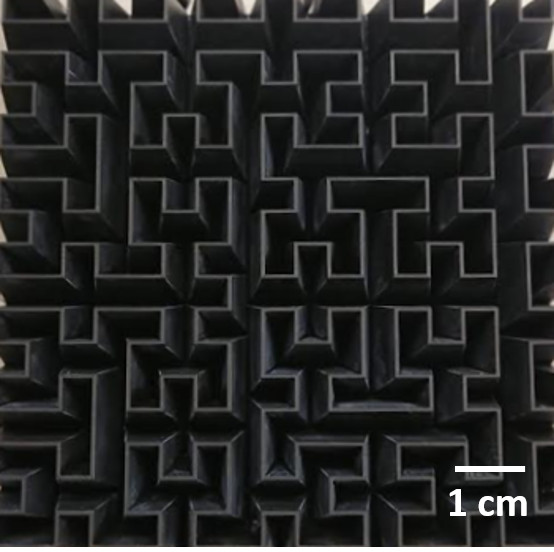}
	\caption{Top-view photographic image of a stereolithographically fabricated, polymethacrylate Hilbert curve absorber sample. A true to form result of the fabricated absorber sample can be recognized. The surface roughness of the absorber is substantially smaller than what had been observed from an absorber fabricated using fused filament deposition \cite{Petroff20193D}.}
	\label{fig_abs}
\end{figure}

Two polymethacrylate samples, a THz absorber and a bulk reference sample, were fabricated using a commercial stereolithography system (Form 2, Formlabs Inc.). For both samples a THz-transparent polymethacrylate \cite{Park2019}, which is compatible with the employed stereolithography system (black, Formlabs Inc.), was used. The THz absorber was designed based on the geometry suggested in Ref.~\cite{Petroff20193D} as shown in Fig.~\ref{fig_curve}. A triangular cross section is extruded along a fractal, space filling Hilbert curve. The Hilbert curve used here is a type of space-filling curves, which are curves that pass through every point of a two-dimensional region with positive Jordan measure \cite{Sagan1994}. The absorber structure was modeled using commercial 3D CAD design software (SolidWorks, Dassault Syst\`emes), by extruding a wedge along the Hilbert curve path as shown in Fig.~\ref{fig_3D}. The dimensions are specified in Fig.~\ref{fig_curve} and \ref{fig_3D}. A top view of the printed absorber sample is shown in Fig.~\ref{fig_abs}. 

The employed stereolithography system uses an inverted bottom-to-top 
fabrication approach including a computer-generated support structure and allows the sample to be arbitrarily oriented during the fabrication process \cite{Taormina2018}. This is crucial here, because the absorber fabrication requires an orientation where the absorber base plane is aligned at an angle of 45$^{\circ}$ with respect to the polymerization plane. This prevents the accumulation of un-polymerized monomer in the void space of the triangular absorber cross-section (Fig.~\ref{fig_3D}).

In addition to the absorber, a bulk reference sample with plane parallel interfaces and a nominal thickness of 8.5~mm was fabricated. The thickness of the reference sample corresponds to the effective thickness of the polymethacrylate present in the absorber sample. Comparing the transmission data obtained from absorber and reference samples thus allows the distinction between the absorption due to the geometry of the absorber and the absorption originating within the polymethacrylate.

After the polymerization process, both samples were immersed in isopropanol for 3~min and then rinsed for an additional 20~min to remove the un-polymerized resin. After the rinsing process, the samples were placed in a precision oven at 60~$^{\circ}$C for 5~min in order to evaporate any excess isopropanol. After the heating process, the samples were placed in a UV oven and were cured for 20~min to ensure complete polymerization.

\subsection{Data Acquisition and Analysis}
Transmission measurements were carried out for the fabricated absorber and the bulk reference sample at normal incidence over the spectral range from 82 to 125~GHz. An electronically controlled signal generator (Synthesizer, Virginia Diodes Inc.) operating in a range from 8 to 20~GHz was used in tandem with a signal generator extension module (Virginia Diodes Inc.) as a source.  

The output signal from the source was collimated with a 60~mm focal length lens before interacting with the sample. The transmitted signal was focused using a second 60~mm focal length lens, and was measured with a broadband power meter (PM3, Erickson Instruments). The source and detector were controlled by LabView VI for automated data acquisition.

\section{Result and Discussion}
\begin{figure}[!t]
	\centering
	\includegraphics[width=0.9\linewidth, trim=0 25 0 50,clip]{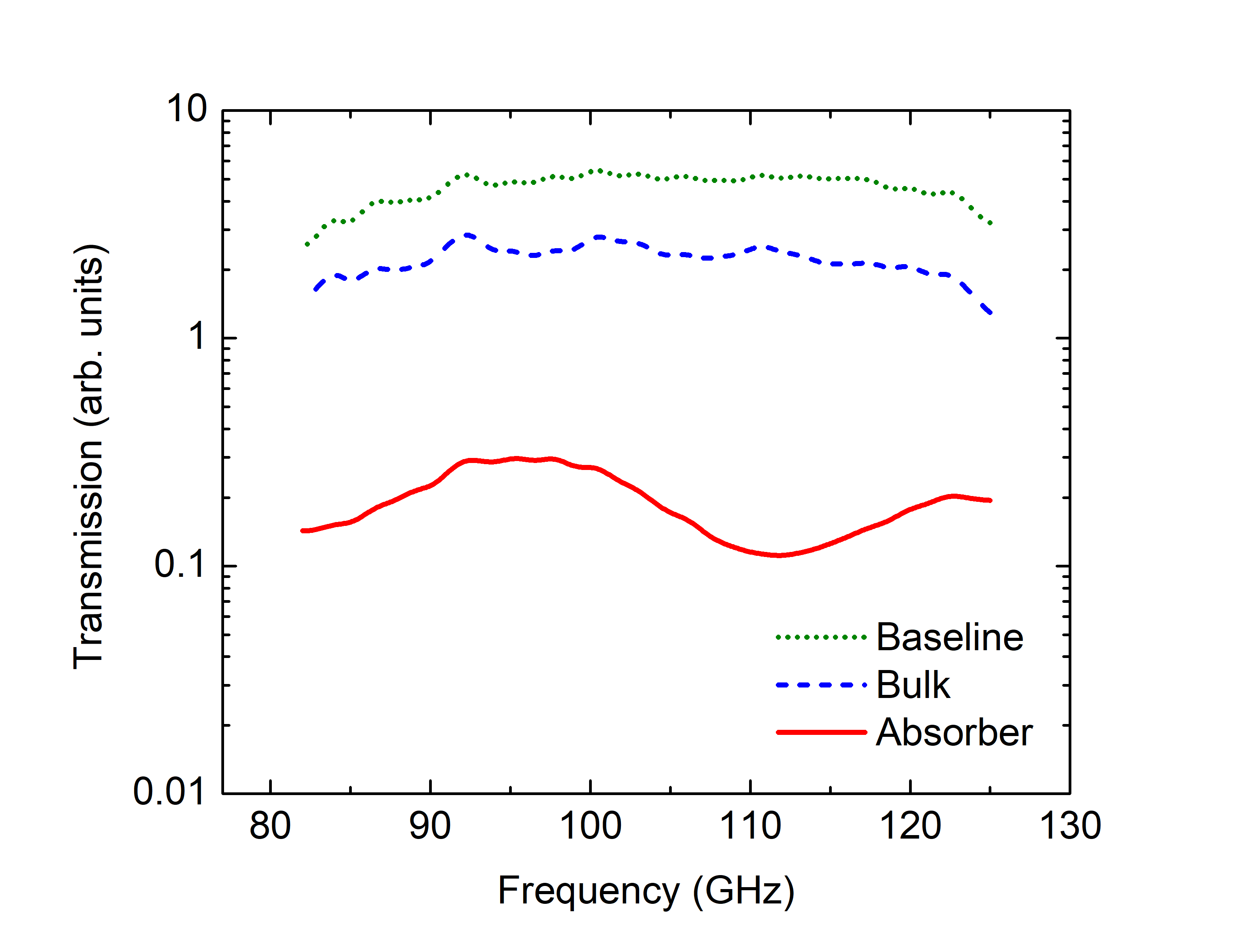}
	\caption{Experimental data of baseline measured without any sample (green dotted line), transmission through the bulk reference sample (blue dashed line), and transmission through the THz wave absorber (red solid line).}
	\label{fig_data}
\end{figure}

Figure~\ref{fig_data} illustrates the experimental transmission data from the reference bulk sample (blue dashed line) and the absorber sample (red solid line) along with the source baseline (green dotted line). The reference sample is attenuating the transmitted signal by approximately 60\% over the entire spectral range, which is corresponding to the THz ellipsometric measurements carried out in our previous work \cite{Park2019}. For the absorber sample, the transmitted signal is reduced substantially stronger and only approximately 1\% of the frequency-averaged incident signal is collected by the detector. 

While the base intensity and the transmitted signal obtained from the reference sample are relatively flat over the investigated spectral range, the transmitted signal obtained through the absorber structure shows a distinct oscillatory behavior. At the peak intensity at 95~GHz the fabricated absorber attenuates approximately 99\% of the incident THz radiation. At the minimum intensity at 110~GHz the absorption is even stronger and 99.5\% of the incident THz radiation is absorbed. While the cause of the oscillatory behavior of the transmitted signal observed from the absorber sample is currently under investigation, we tentatively assign it to an interference within the absorber walls, which have plane parallel interfaces. This interpretation is supported by calculations using the refractive index of the employed polymethacrylate while assuming a wall thickness on the order of 1~mm \cite{Park2019}.

\section{Conclusion}

A broadband THz wave absorber was stereolithographically fabricated using a polymethacrylate.
The transmissive behavior of this absorber was compared to a reference sample with the same effective material thickness. While the reference sample transmits approximately 40\% of the incident light in the spectral range from 82 to 125~GHz, the absorber attenuates strongly. Only 1\% of the frequency-averaged incident signal is transmitted. The strongest attenuation is found at 110~GHz where 99.5\% of the incident THz radiation is absorbed. 

The absorber studied here was designed using a space-filling Hilbert curve along which a triangular cross section is extruded. A number of alternative space-filling fractal curves exist, including Peano, Moore, and Gosper curves, for instance \cite{Gardner1976, Bader2013, Sagan1994}. In comparison to the Hilbert curve, some of these variants have non-orthorgonal unit cells and could offer advantages for the fabrication and automated tool path generation \cite{Cox1994}. In addition, surfaces fabricated using such alternative curves might offer different polarization responses. However, absorber structures based on these alternative curves have not been explored yet. 

The results demonstrated that effective THz absorbers can be fabricated using polymethacrylate-based stereolithography. Our observations further corroborate the strong absorbing behavior deduced from the reflection measurements reported by Petroff \textit{et al.} on absorber structures manufactured using fused filament deposition \cite{Petroff20193D}.

In conclusion, it is found here that stereolithography can provide an effective fabrication approach for THz absorber structures. In comparison with other additive manufacturing methods, stereolithography provides superior spatial resolution and surface finish. However, enclosed void spaces can be challenging to fabricate using this technique as un-polymerized monomer has to be effectively removed during and after the polymerization process. This fabrication challenge is resolved here by fabricating the samples at an oblique angle with respect to the polymerization plane. This leads to monomer drainage during the polymerization process. Remaining monomer can be effectively removed using standard post process, rinsing procedures in suitable solvents. 

\section*{Acknowledgment}

\noindent SP, YL, and TH would like to acknowledge the valuable discussions with Susanne Lee and Erin Sharma within the NSF I/UCRC for Metamaterials. The authors are grateful for support from the National Science Foundation (1624572) within the I/UCRC Center for Metamaterials, the Swedish Agency for Innovation Systems (2014-04712), and the Department of Physics and Optical Science of the University of North Carolina at Charlotte.


\begin{thebibliography}{10}
	\providecommand{\url}[1]{#1}
	\csname url@samestyle\endcsname
	\providecommand{\newblock}{\relax}
	\providecommand{\bibinfo}[2]{#2}
	\providecommand{\BIBentrySTDinterwordspacing}{\spaceskip=0pt\relax}
	\providecommand{\BIBentryALTinterwordstretchfactor}{4}
	\providecommand{\BIBentryALTinterwordspacing}{\spaceskip=\fontdimen2\font plus
		\BIBentryALTinterwordstretchfactor\fontdimen3\font minus
		\fontdimen4\font\relax}
	\providecommand{\BIBforeignlanguage}[2]{{%
			\expandafter\ifx\csname l@#1\endcsname\relax
			\typeout{** WARNING: IEEEtran.bst: No hyphenation pattern has been}%
			\typeout{** loaded for the language `#1'. Using the pattern for}%
			\typeout{** the default language instead.}%
			\else
			\language=\csname l@#1\endcsname
			\fi
			#2}}
	\providecommand{\BIBdecl}{\relax}
	\BIBdecl
	
\bibitem{zhou2016additive}
F.~Zhou, W.~Cao, B.~Dong, T.~Reissman, W.~Zhang, and C.~Sun, ``Additive
{M}anufacturing of a 3D Terahertz Gradient-Refractive Index
{L}ens,'' \emph{Adv. Opt. Mater.}, vol.~4, no.~7, pp. 1034--1040, 2016.

\bibitem{squires20153d}
A.~Squires, E.~Constable, and R.~A. Lewis, ``3D Printed Terahertz Diffraction
Gratings and Lenses,'' \emph{J. Infrared, Millimeter, Terahertz Waves},
vol.~36, pp. 72--80, 2015.

\bibitem{Weidenbach:16}
M.~Weidenbach, D.~Jahn, A.~Rehn, S.~F. Busch, F.~Beltr\'{a}n-Mej\'{i}a, J.~C.
Balzer, and M.~Koch, ``3D Printed Dielectric Rectangular Waveguides,
Splitters and Couplers for 120 GHz,'' \emph{Opt. Express}, vol.~24, no.~25,
pp. 28\,968--28\,976, Dec 2016.

\bibitem{Yan1996}
\BIBentryALTinterwordspacing
X.~Yan and P.~Gu, ``A Review of Rapid Prototyping Technologies and Systems,''
\emph{Computer-Aided Design}, vol.~28, no.~4, pp. 307 -- 318, 1996.
\BIBentrySTDinterwordspacing

\bibitem{NGO2018172}
\BIBentryALTinterwordspacing
T.~D. Ngo, A.~Kashani, G.~Imbalzano, K.~T. Nguyen, and D.~Hui, ``Additive
Manufacturing (3D Printing): A Review of Materials, Methods, Applications and
Challenges,'' \emph{Composites Part B: Engineering}, vol. 143, pp. 172 --
196, 2018. 
\BIBentrySTDinterwordspacing

\bibitem{Shallan2014chem}
A.~I. Shallan, P.~Smejkal, M.~Corban, R.~M. Guijt, and M.~C. Breadmore,
``Cost-Effective Three-Dimensional Printing of Visibly Transparent Microchips
Within Minutes,'' \emph{Analytical chemistry}, 2014.

\bibitem{Fullager2019}
\BIBentryALTinterwordspacing
D.~B. Fullager, S.~Park, C.~Hovis, Y.~Li, J.~Reese, E.~Sharma, S.~Lee,
C.~Evans, G.~D. Boreman, and T.~Hofmann, ``Metalized Poly-methacrylate
Off-Axis Parabolic Mirrors for Terahertz Imaging Fabricated by Additive
Manufacturing,'' \emph{J. Infrared, Millimeter, Terahertz Waves}, vol.~40,
no.~3, pp. 269--275, Mar 2019.
\BIBentrySTDinterwordspacing

\bibitem{colla2019}
J.~A. Colla, R.~E.~M. Vickers, M.~Nancarrow, and R.~A. Lewis, ``3D Printing
Metallised Plastics as Terahertz Reflectors,'' \emph{J. Infrared, Millimeter,	Terahertz Waves}, vol.~40, pp. 1--11, 2019.

\bibitem{Park2019}
S.~Park, Y.~Li, D.~Fullager, S.~Sch\"{o}che, C.~Herzinger, G.~Boreman, and
T.~Hofmann, ``Terahertz to Mid-Infrared Dielectric Properties of
Polymethacrylates for Stereolithographic Single Layer Assembly,'' \emph{J.
	Infrared, Millimeter, Terahertz Waves}, vol.~40, no.~9, pp.~971--979, Sep 2019.

\bibitem{Petroff20193D}
\BIBentryALTinterwordspacing
M.~Petroff, J.~Appel, K.~Rostem, C.~L. Bennett, J.~Eimer, T.~Marriage,
J.~Ramirez, and E.~J. Wollack, ``A 3D-Printed Broadband Millimeter Wave
Absorber,'' \emph{Rev. Sci. Instrum.}, vol.~90, no.~2, p. 024701, 2019.
\BIBentrySTDinterwordspacing

\bibitem{Sagan1994}
H.~Sagan, \emph{Hilbert's Space-Filling Curve}. Springer, New York, NY, 1994.

\bibitem{Taormina2018}
G.~Taormina, C.~Sciancalepore, M.~Messori, and F.~Bondioli, ``3d printing
processes for photocurable polymeric materials: technologies, materials, and
future trends,'' \emph{Journal of applied biomaterials \& functional materials}, vol.~16, no.~3, pp. 151--160, 2018.

\bibitem{Gardner1976}
M.~Gardner, ``Mathematical Games,'' \emph{Sci. Amer.}, vol. 235, pp. 206--211, 1976.

\bibitem{Bader2013}
M.~Bader, \emph{Grammar-Based Description of Space-Filling Curves}.\hskip 1em
plus 0.5em minus 0.4em\relax Berlin, Heidelberg: Springer Berlin Heidelberg,
2013, pp. 31--45.

\bibitem{Cox1994}
J.~J. Cox, Y.~Takezaki, H.~R. Ferguson, K.~E. Kohkonen, and E.~L. Mulkay,
``Space-Filling Curves in Tool-Path Applications,'' \emph{Computer-Aided
	Design}, vol.~26, no.~3, pp. 215--224, 1994.
	
\end{thebibliography}


\end{document}